\documentstyle[prl,aps,multicol]{revtex}
\input epsf

\begin{document}

\title{Nonlocal Aspects of a Quantum Wave}

\author{Y. Aharonov$^{a,b}$,  and L. Vaidman$^a$}

\maketitle
\vspace{.4cm}
 
\centerline{$^a$ School of Physics and Astronomy}
\centerline{Raymond and Beverly Sackler Faculty of Exact Sciences}
\centerline{Tel-Aviv University, Tel-Aviv 69978, Israel}
\vskip .2cm
\centerline{$^b$ Physics Department, University of South Carolina}
\centerline{Columbia, South Carolina 29208, USA}

\date{}

\vspace{.4cm}
\begin{abstract}
  Various aspects of nonlocality of a quantum wave are discussed. In
  particular, the question of the possibility of extracting
  information about the relative phase in a quantum wave is analyzed.
  It is argued that there is a profound difference in the nonlocal
  properties of the quantum wave between fermion and boson particles.
  The phase of the boson quantum state can be found from correlations
  between results of measurements in separate regions. These
  correlations are identical to the Einstein-Podolsky-Rosen (EPR)
  correlations between two entangled systems. An ensemble of results
  of measurements performed on fermion quantum waves does not exhibit
  the EPR correlations and the relative phase of fermion quantum waves
  cannot be found from these results. The existence of a physical
  variable (the relative phase) which cannot be measured locally is
  the nonlocality aspect of the quantum wave of a fermion.

\end{abstract}
\vspace {.5cm}

\begin{multicols}{2}

\section{INTRODUCTION}
\label{int}

There are literally thousands of papers about nonlocality in quantum
theory. However,  there are still some aspects of
nonlocality which have not been fully explored and the connection between
various aspects have not been clarified. In this paper we will analyze
 particular nonlocal aspects which are different for quantum waves of
bosons and fermions. This is a development of  ideas originated in
the works of one of us  \cite{A-non,AA}. In order to put these
nonlocality aspects in the proper perspective 
 we will give a brief review of other
aspects of nonlocality of quantum theory.

 An important nonlocality aspect which will not be 
discussed in this paper is related to the concept of {\em  nonlocal
  variables}.  Measurements of  nonlocal variables cannot be reduced to measurements of
local variables \cite{AAV86}. Probably the simplest example of a nonlocal variable is the sum of spin
components of  two separated spin-$1\over2$ particles, 
$\sigma_{Az} + \sigma_{Bz}$. According to the postulates of the
quantum theory, if a system is in an eigenstate of a measured variable,
ideal measurement of this variable should not alter this eigenstate. For
example, the singlet state of the two spins, frequently named the
Einstein-Podolsky-Rosen (EPR) state,
 \begin{equation}
\label{sing}
 |\Psi\rangle_{EPR} = {1\over {\sqrt 2}}(  |{\uparrow}\rangle_A
|{\downarrow}\rangle_B -   |{\downarrow}\rangle_A
|{\uparrow}\rangle_B),  
\end{equation}
is an  eigenstate of the operator $\sigma_{Az} + \sigma_{Bz}$ with an
eigenvalue $0$. Thus, measurement of $\sigma_{Az} + \sigma_{Bz}$ must
leave state (\ref{sing}) unchanged. Note that measurements of local variables,
 $\sigma_{Az}$ and $ \sigma_{Bz}$, invariably
change the state.

Some of the eigenstates of the nonlocal variable $\sigma_{Az} +
\sigma_{Bz}$ are entangled states. It is interesting that there are
nonlocal variables which have only product-state eigenstates. They are
nonlocal in the  sense of impossibility of their measurement using only
local measurements in the space-time regions  A and B \cite{PV}.
Moreover, recently \cite{nwe} there have been found nonlocal variables with
product-state eigenstates which cannot be measured even when
measurements are performed at different times in space locations A and B and
unlimited classical communication between the sites is allowed.

Very important  nonlocal variables are {\em modular
  variables} \cite{mod}. Many surprising effects related to evolution of
spatially separated systems can be effectively analyzed using them. The dynamical equations of modular variables are nonlocal.

However, the nonlocality issues related to nonlocal variables are
mingled: it is not easy to separate which part of nonlocality in the
dynamics is due to intrinsic nonlocality of the quantum world and
which part is due to the nonlocality introduced by the definition of
the variable. In this paper we limit ourselves to the analysis of
relations between result of measurements of local variables.

The plan of the paper is as follows. In Section \ref{frame0} we
introduce the basic framework of our analysis. In Sections
\ref{colla}-\ref{ABnon} we discuss three types of nonlocality. This
discussion provides the frame of reference for the analysis of
nonlocality. In Section \ref{frame} we give a more detailed explanation
of the framework. Following this preparatory introduction,  we analyze the
nonlocality of the boson quantum wave in Section \ref{spnon} and the
nonlocality of the fermion quantum wave in Section \ref{ferm}. Section
\ref{disc} is devoted to an apparent causality paradox arising
from nonlocality of the boson wave. In Section \ref{swap} we discuss a
related issue of collective measurement which is relevant mostly for
the fermion quantum wave. Finally, in Section \ref{conc} we summarize
the main results of the paper.

\section{THE FRAMEWORK OF THE ANALYSIS}
\label{frame0}

The formalism of non-relativistic quantum theory allows introducing
arbitrary Hamiltonians, in particular, Hamiltonians corresponding to
nonlocal interactions.  However, such interactions have not been
observed in experiments.  In the framework of our analysis of
nonlocality we will assume that the Hamiltonian describes only local
interactions. This is a basic assumption of our analysis.

Any wave  in space is, in some sense,  a
nonlocal object.  A classical wave, however, can be considered as a
collection of local properties. What makes the  quantum wave genuinely
nonlocal is that it cannot be reduced to a collection of local
properties.  In order to analyze
this aspect of a quantum wave we will concentrate on  a particular
simple case: a quantum wave which is an equal-weights superposition of
two localized wave packets in two separate locations:
 \begin{equation}
\label{qw}
|\Psi \rangle = 
{1\over \sqrt 2} (    | a\rangle +
~e^{i\phi} | b \rangle).
\end{equation}
 We will analyze various simultaneous (in a particular
 Lorentz frame)
measurements performed in these two locations; see Fig. 1.  
We will denote by A and B the space-time regions of these measurements.
The wave packet  $| a\rangle$ is localized inside the spatial region of A
and the wave packet  $| b\rangle$  is localized inside the spatial region of B.

\begin{center} \leavevmode \epsfbox{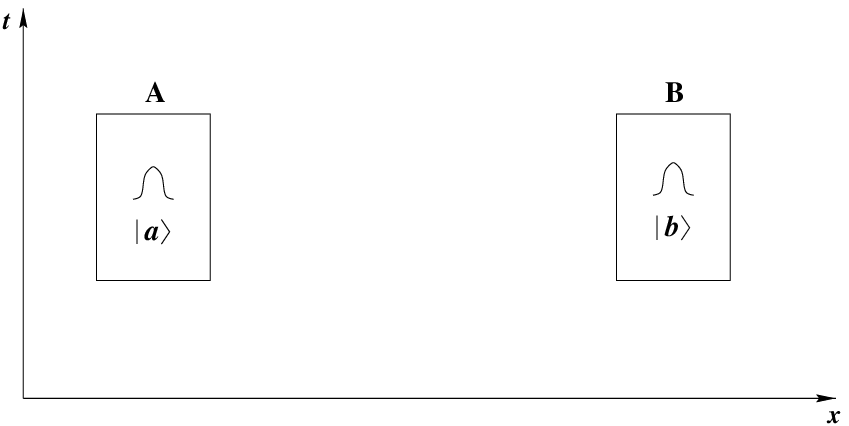} \end{center}

\noindent 
{\small {\bf Fig. 1.} Space-time diagram of the measurements performed
  on the quantum wave (\ref{qw}).}

\vskip .4cm

In this paper we will show that there is a profound difference in the
nonlocal properties of the quantum wave of the form (\ref{qw}) between
fermion and boson particles.  The boson state  leads to
statistical correlations between results of measurements in A and in B
that {\it cannot} be explained by local classical physics.  The fermion
state  does not lead to such correlations but it has a
different nonlocality aspect. The fermion quantum state cannot be measured
using only local measurements in A and B even if we are given an
ensemble of results of measurements performed on identical particles
in the state (\ref{qw}).  In particular, the relative phase $\phi$ of
the fermion state does not lead to locally measurable effects. This
phase has a physical meaning: it influences the result of interference
experiments in which the parts of the quantum state in A and in B are
brought together. Existence of a physical quantity which does not
manifest itself through local measurements is the nonlocality aspect of a
fermion wave. In contrast, a boson state can be found  from the
ensemble of results of local measurements: it can be identified  from the
nonlocal correlations mentioned above.

\section{NONLOCALITY OF THE COLLAPSE OF A QUANTUM STATES}
\label{colla}

In a situation in which a particle (boson or fermion) is described by
the state (\ref{qw}), each region, A or B separately, cannot be
described by a pure quantum state. By introducing the {\em vacuum}
states $|0\rangle_A$ and  $|0\rangle_B$ which describe the regions A
and B without the particle, we can rewrite the state  (\ref{qw}) in
the following form
 \begin{equation}
\label{qw2}
|\Psi \rangle = 
{1\over \sqrt 2}\ (  | 1\rangle_A |0\rangle_B +
~e^{i\phi} \,| 0 \rangle_A |1\rangle_B ) ,
\end{equation}
where $ | 1\rangle_A\equiv |a\rangle$ and $ | 1\rangle_B\equiv
|b\rangle$. This form allows us to write down the complete quantum
description of region B (as well as region A)  by means of the density matrix
    \begin{equation}
      \label{rho0}
    \rho_0 =  \left( \begin{array}{cc} 1 \over 2 &0\\
0 & 1\over 2
\end{array}
\right).
    \end{equation}
In the framework of standard  quantum theory, a measurement
instantaneously collapses the quantum state of a system. Thus, an action in A can
change the density matrix in B.
After a measurement of the projection  operator in A, i.e., after
observing  whether 
the particle is in A, the density
matrix in B is  changed instantaneously to the density matrix
of one of the pure states:
    \begin{equation}
      \label{rho1}
    \ \left( \begin{array}{cc} 1  &0\\
0 & 0
\end{array}\right) ~~ {\rm or} ~~
 \left( \begin{array}{cc} 0 &0\\
0 & 1
\end{array}
\right),
    \end{equation}
in anti-correlation to the corresponding  density matrices in A.

According to the collapse interpretation, the measurement in A changes
the state of affairs in B. Before the action in A the outcome of a
possible measurement of the projection operator in B was undetermined
not only to the observer in B, but to all. Nothing in nature could give an
indication about the outcome of the experiment. The outcome is
genuinely random with  probability  $1\over 2$ both for finding
 region B empty and for finding the particle there. After the
measurement in A, the observer in B still does not know the outcome,
but  nature (in particular, the observer in A) has this
information: the probabilities for the results of the measurement in B change
to either 1 and 0 or to 0 and 1 according to the outcome in A.

There is no other example in physics in which a local action 
changes the state of affairs in a space-like separated region. Thus, this aspect of nonlocality provides an argument in favor of adopting
one of the interpretations which does not have the collapse of a
quantum state. We now briefly  describe  these interpretations.
    
According to the pragmatic approach \cite{Peres}, quantum theory is
limited to providing a recipe for predicting probabilities in quantum
experiments, i.e.  frequencies of the outcomes in the experiments.
 In this approach the density matrix is a statistical concept.
An observer in B, who does not know which outcome is obtained in A,
considers the mixture of the two possibilities (\ref{rho1}) as
described by the statistical density matrix $\rho_0$ even after the
measurement in A.
    
The causal interpretation of Bohm \cite{Bohm52,Bell81} has no collapse
and therefore it lacks the nonlocality aspect of instantaneous change
of a quantum state.  The result of the measurement of projection
operators on region B is predetermined by a ``Bohmian position'' and,
therefore, the measurement in A changes nothing in B. For a single
particle, Bohmian theory is a local hidden variable theory which
completes quantum mechanics without contradicting statistical
predictions of the latter.  However, for systems consisting of more
than one particle, the evolution of ``Bohmian positions'' of the
particles is nonlocal.  The Bohmian theory is nonlocal in a robust
sense: action in A can change the outcome in B. For example, consider
the EPR state of two spin-$1\over2$ particles (\ref{sing}).  Consider
Bohmian positions which are such that if a particular $\sigma_z$
measurement is performed on either particle, it must yield $\sigma_z
=1$. However, if these $\sigma_z$ measurements are performed on both
particles, the results will be different: the earlier
measurement of $\sigma_z$ in A will change the outcome of the
consequent measurement in B to $\sigma_z =-1$.  The details of this
example are given in Ref.  \cite{V-ghz}.

The non-collapse interpretation which one of us (L.V.) finds most
appealing \cite{V-mwi} is the many-worlds interpretation (MWI)
\cite{mwi}. In the physical universe, due to  the measurement in A, the
quantum state of the two particles and the measuring device in A
changes in the following way:
 \begin{eqnarray}
\label{split}
\nonumber
 {1\over {\sqrt 2}}(  |1\rangle_A
|0\rangle_B -   |0\rangle_A
|1\rangle_B)  |{\rm ready}\rangle_{{MD_A}} \rightarrow ~~~~~~~~~~~~~~~~~
\\ 
  {1\over {\sqrt 2}}(  |1\rangle_A |0\rangle_B |{\rm click}\rangle_{{MD_A}}
 -   |0\rangle_A |1\rangle_B |{\rm no~click}\rangle_{{MD_A}})  ,
\end{eqnarray}
but the density matrix in B is still $\rho_0$. Note, that relative to
an observer in A, who belongs to a world with a particular reading of
the measuring device, the density matrix of the particle in B is that
of one of the  pure states (\ref{rho1}). Only from the point of view of an
external observer, who is not correlated to a particular outcome in A,
the density matrix in A is unchanged.

If we now add  an observer in B who measures the
projection operator there, then in A there is a mixture of two worlds
with and without the particle in A and, similarly, in B there is a
mixture of two worlds with and without the particle in B. These
mixtures were created locally by the decisions of the observers to
make these particular measurements.  What remains nonlocal in this
picture are the ``worlds'': the observer in A who found the
particle, in his travel to B, will meet there the observer  that has not
found the particle, and
vice versa in the other world.

One of us (Y.A.) strongly prefers an interpretation which does not
require a multitude of worlds. The two-state vector formalism of quantum
theory \cite{ABL,AV} allows covariant description of the collapse.
This picture suggests radical change in the concept of time which will
avoid statements made above such as: ``According to the collapse
interpretation, the measurement in A changes the state of affairs in
B.''  These ideas  will  be presented elsewhere.

\section{NONLOCALITY OF CORRELATIONS}
\label{corr}

In the framework of standard quantum theory the (anti)correlations
between finding particles in the two regions A and B described above
are nonlocal in the sense that the theory does not yield a causal
explanation for them. The complete quantum description does not
specify the results of measurements and it does not yield a local causal
explanation for this correlations. One might imagine that the quantum
theory can be completed by a deeper theory which will provide a local
causal explanation for the results of measurements. In fact, the
Bohmian theory mentioned above provides a local explanation for the
anti-correlations in finding the particle in the regions A and B, but
for some other experiments performed in these space-like separated
regions it is impossible to find a local hidden variable theory. In
particular, statistics of the results of spin measurements performed
on two separated spin-$1\over 2$ particles in a singlet state (the
setup for which Bohmian theory is not local) cannot be explained by
a local hidden variables theory.  This is the content of the celebrated
Bell-inequalities paper \cite{Bell64}.

There are numerous proofs that quantum correlations cannot have local
causes.   We  present here one more
argument of this kind inspired by the work of Mermin
\cite{Mer}. However, the reader  can choose  any other proof of this
statement in order to proceed with the line of argumentation of this
paper.

The argument presented here assumes the principle of counterfactual
definiteness \cite{Read}, i.e., that in any physical situation the
result of any experiment which can be performed has a definite value.
We will analyze again the EPR state (\ref{sing}). Consider
measurements of the spin components in  $N+1$ directions for the
particle in A and in  $N$ different directions for the particle in B. These
directions are in the $\hat x -\hat z$ plane and they are
characterized by the angle $\theta_i$ with respect to the $\hat z$
axis,
\begin{equation}
  \label{theta}
\theta_i \equiv {{i\pi}\over{2N}}, ~~~~i =0,1, ...2N.
\end{equation}
Note that the measurement in the direction $\theta_{2N}(=\pi)$ is physically
equivalent to the measurement in the direction $\theta_0(=0)$, but the
result has to be multiplied by $-1$, i.e.,  $\sigma (\pi) = - \sigma (0)$.

Spin measurement of one particle in a given direction (effectively)
collapses the spin state of the other particle to the opposite
direction and, therefore, quantum theory predicts the same probability
for all the following relations between the results of measurements,
if performed:
\begin{eqnarray}
  \label{resu}
  \sigma_A  (  \theta_{2n}) =  -\sigma_B ( \theta_{2n+1}) ,\\
  \label{resu1}
  \sigma_A  ( \theta_{2n+2}) =  -\sigma_B ( \theta_{2n+1}) , 
\end{eqnarray} 
where $n =0,1,...,N-1$.  
The probability is
\begin{equation}
  \label{prob}
  p = \cos ^2 ({ {\theta_{i+ 1} - \theta_{i}}\over 2}) =
 \cos ^2 ({\pi \over {4N}}).
\end{equation}

From the principle of counterfactual definiteness and the locality
assumption, according to which local measurements yield the same
outcomes independently of what has been measured in the other
location, it follows that identical expressions in the equations (\ref{resu})
and (\ref{resu1}) must correspond to equal values. Thus, we can use
all these $2N$ equations together. The correctness of all the equation
leads to a contradiction. Indeed, we obtain $ \sigma_A ( \theta_{0}) =
\sigma_A ( \theta_{2N})$ contrary to the fact that these expressions 
represent the same measurement in opposite  directions:
 $\sigma_A (0) = - \sigma_A (\pi)$. Therefore, at least one out of $2N$
equations (\ref{resu}) and (\ref{resu1}) {\em must} fail to be
satisfied.  On the other hand, irrespective of what correlations (compatible with quantum mechanics) follow from a
hidden variable theory,
the probability that at least one of these relations fails  to be
satisfied cannot be
more than the probability that one fails multiplied by the number of
equations:
\begin{equation}
  \label{probN}
 {\rm prob(fail)}\leq 2N (1-p) = 2N (1- \cos ^2 ({\pi \over {4N}})) .
\end{equation}
This expression, however, is smaller than 1 even for $N=2$ and for
large $N$ it goes to zero as ${\pi ^2\over {8N}}$.

Recently, Greenberger, Horne and Zeilinger (GHZ) \cite{GHZ} have found
an   even more robust example (improved by Mermin \cite{Mermin}) of
such nonlocality. While in our example  we have several relations which have to be
true according to quantum theory with high probability, in spite of the
fact that they all  cannot be true,  in the GHZ example we have four
relations which must be true with probability 1, but, nevertheless,
they cannot all be true together. However, in the GHZ(Mermin) example we have
to consider  three,
instead of two, space-like separated regions. 

Note that the Bell and the GHZ arguments do not hold  without  the
principle of counterfactual definiteness, i.e., it is not applicable
in the framework of the many-worlds interpretation in which, in general,
quantum measurements do not possess single outcomes.

\vskip .5cm
{\em }

\section{THE AHARONOV-BOHM TYPE NONLOCALITY}
\label{ABnon}

Another  nonlocality aspect of quantum theory is related to
the Aharonov-Bohm (AB) effect. The effect has a topological basis.
The wave-function of a particle enters two space regions tracing out
trajectories in space-time which start and end together.  An
interference pattern which depends upon a field is observed in spite
of the fact that locally, inside these regions, it is impossible to
make measurements which can specify the result of the interference
experiment. The main  aspect of the effect is that it exists
even when there is no field inside the regions during the whole time
of the experiment.

In this paper we consider measurements in two space-time regions.
This is different from the AB effect for which a closed trajectory in
space is required.  What is relevant to our discussion is the feature
of a particle inside the two space-time regions A and B which will
eventually be manifested in the results of the interference
experiment. The AB nonlocality is the existence of a physical property
(a property which has observable consequences) which does not have any
manifestation in local measurements.

A simple example is a particle wave-packet which splits into a
superposition of two wave packets (\ref{qw}) and later brought back
again to the same region for an interference experiment. This can be
achieved in a one dimensional model of a wave packet arriving at a
barrier at time $t_0$; see Fig. 2a. The barrier is such that the
particle has the probability $1\over 2$ to pass through and the
probability $1\over 2$ to be reflected. Two reflecting walls at equal
distance from the barrier return the two wave packets back to the
barrier at the same time and the result of the interference experiment
is observed by finding the particle on the left or on the right side
of the barrier at a later time.  The time-dependent (scalar) AB effect
is obtained by changing the relative potential between the two parts
of the wave during the time they are separated. For a charged particle
this can be achieved by moving two large oppositely charged parallel
plates located between the wave packets; see Fig. 2b. The two plates
are placed originally one on top of the other, i.e., there is no
charge distribution and, therefore, there is no electric field
anywhere. The plates are then moved a short distance apart and then
they are brought back. We will call such an operation ``opening a
condenser''.

\begin{center} \leavevmode \epsfbox{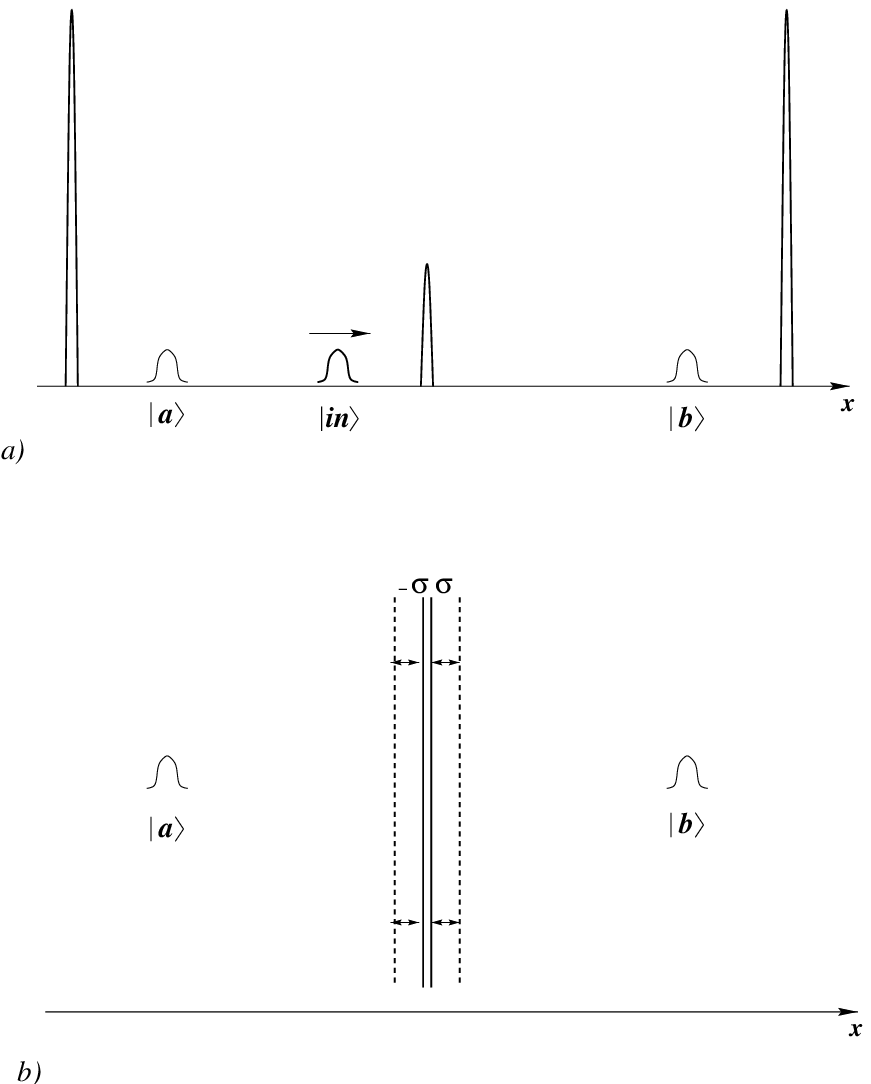} \end{center}

\noindent 
{\small {\bf Fig. 2.} Scalar Aharonov-Bohm effect interference experiment. {\em
    a}). One-dimensional interference experiment. The particle in the
  wave packet $|in\rangle$ splits at the barrier into a superposition of
  the two wave packets, $| a\rangle$ and $| b\rangle$, which are
  reflected from the walls and reunited to interfere at the
  barrier. {\em b}). Parallel-plate condenser with charged plates,
  originally one on top of the other, is opened (by moving the plates
  apart) for a short time while the wave packets  $| a\rangle$ 
and   $| b\rangle$ are far apart. This operation introduces change in
the electric potentials between the locations of  $| a\rangle$ 
and   $| b\rangle$  which generates the AB phase.}

\vskip .4cm

A naive answer to the question, ``What is the nonlocal feature of the
two regions A and B?'' (the feature of the two parts of the wave after
they are separated) would be the quantum phase $\phi$ appearing in
the equation (\ref{qw}).  Indeed, we will argue, discussing fermions
in Section \ref{ferm}, that in certain circumstances the quantum phase
is a nonlocal feature in the sense that it cannot be found through
local experiments in A and in B.  However, the statement is not
correct for bosons.  Moreover, the phase is not a gauge invariant
concept.  The physical effect of interference is of course gauge
invariant since it is a topological property of the whole
trajectory.  Still, there is a property of the system in A and B which
specifies the final outcome of the interference experiment given fixed
circumstances. The quantum phase does characterize this property
provided we are careful enough to fix the gauge in the problem.

This and preceding sections described nonlocality aspects which are
very different: here we discuss an observable property of a system in two
locations which does not have any local manifestation,  while in the
previous section we discussed results of local measurements which do
not allow local-cause explanation. It is possible to perform
analysis of these nonlocalities using different terms, such as {\em local
  action}, {\em separability}, etc. Then the differences between the
nonlocalities discussed in the two sections might  not be as sharp as stated above
\cite{He}. However, such analysis strongly depends on the
interpretation of quantum theory and is less helpful for the purpose
of the present paper.

\section{THE DETAILED FRAMEWORK OF THE ANALYSIS}
\label{frame}

Our goal in this paper is to perform an analysis of nonlocal aspects
of the quantum state (\ref{qw}). The main question is: ``What are the
physical consequences of the presence of this quantum wave in the
space-time regions A and B?''  One of the  questions is: ``Can we
find the quantum phase $\phi$ through local measurements in A and B?''
In order to be able to make such analysis we have to specify exactly
the meaning of space-time regions A and B. Are the positions of A and
B fixed relative to each other or are they fixed relative to an external
reference frame? Are there fixed directions in A and B such that
measuring devices can be aligned according to them? Is the time in
A and B defined relative to local clocks, or relative to an external
clock?  What are the measuring devices which are available in A and B?
All these questions are relevant. We have to specify what is given in
A and B prior to bringing the quantum wave there in order to
distinguish effects related to the quantum wave from the effects
arising from our preparation and/or definition of the sites A and B.

We make the following assumptions:

(i) There is an external inertial frame which is massive enough so
that it can be considered classical.

(ii) There is no prior entanglement of physical systems between the
sites A and B. The two laboratories in A and B are also massive enough
so that the measurements performed on the quantum wave can be
considered measurements performed with classical apparatuses. However,
for various aspects of our analysis we will have to consider the two
laboratories as quantum systems. We assume that relative to the
external reference frame the two laboratories are initially described by a
product quantum state $|\Psi_A\rangle |\Psi_B\rangle$.

(iii) There is no entanglement between location of the apparatuses in A
and  the wave packet $|a\rangle$ (as well as between location of the
apparatuses in B and the wave packet $|b\rangle$). Instead, the fact that
apparatus A measures  $|a\rangle$ and apparatus B measures  $|b\rangle$ is
achieved via localization relative to the external frame. The measuring
devices and the wave packets  are well
localized at the same place. This can
be expressed in the equations
\begin{eqnarray}
  \label{xx}
  \langle a|\hat x|a \rangle =  \langle \hat x_{MD_{A}} \rangle, \\
  \langle b|\hat x|b \rangle =  \langle \hat x_{MD_{B}} \rangle, 
\end{eqnarray}
where $ x_{MD_{A}}$ ($ x_{MD_{B}}$) is the variable which describes the location of
the interaction region of the measuring devices in A (in B). It is
assumed that the wave packet $|a\rangle$ remains in the space region A
(and $|b\rangle$ remains in B) during the time of measurements.

(iv) Measurements in A and in B are performed by local measuring
devices activated by  {\em local} clocks, say, at
the internal time $\tau_A =\tau_B =0$. The clocks are well
synchronized with the time $t$ of the external (classical) clock:
\begin{equation}
  \label{tt}
  \langle \tau_A(t) \rangle =  \langle \tau_B(t) \rangle = t,
\end{equation}
and the spreads of the clock pointer variables $\Delta \tau_A,~\Delta
\tau_B$ are small during the experiment. Again, as stated in (ii),
there is no entanglement between clocks in A and in B.

The assumptions can be summarized as follows: a measurement in A, the
space-time point relative to an external classical frame, means
a measurement performed by local apparatuses in A triggered by the local
clock. The apparatuses and clocks in A are not entangled with the
apparatuses and the clocks in B. 

Given all apparatuses in A and B, but without the quantum particle
(\ref{qw}), it is impossible to observe the nonlocality of the collapse
described in Section \ref{colla}. Since the quantum state of all
systems (measuring apparatuses, clocks, etc.) is the product state of
a quantum state in spatial location A times a quantum state in spatial
location B, there are no correlations between the results of
measurements in A and in B. This requirement need not be so strong:
the crucial feature is the absence of quantum correlations (following
from entanglement between the systems in A and in B). Here, for
simplicity of the analysis we forbid any initial correlation between
measuring devices in the two sites.

There is a somewhat more complicated situation in relation to the
nonlocality discussed in Section \ref{ABnon}. Clearly there is no
quantum phase which characterizes the devices in A and  B: these
systems are in the product state. But the operational definition of
the AB nonlocality of Section \ref{ABnon} was a feature which cannot
be found through local experiments in A and B, the feature which leads
to observable effects when the systems from locations A and B are
brought together. If we restrict ourselves to measurements using local
measuring devices, then there are many features which cannot be found
locally, for example, the relative orientation of the measuring
devices in A and in B.  The observer in A (or in B) making
measurements using local devices cannot find out his (or her)
orientation. However, if we have other observers in the product state
in regions A and B with well defined {\em known} orientation, they can
measure locally the orientation of the system in A and orientation of
the system in B. The question of  what can and what cannot be measured
from within the system itself is interesting \cite{AR,CR}, but we will
not discuss it here. Here we allow all possible measuring devices
provided that they do not possess entanglement between A and B.

In our discussion we assume that measurements are performed on a single
system. But, for the question of  finding  the phase,
the question  of obtaining non-classical correlations, etc., we
assume that we have an ensemble of experiments on identical single systems.
Collective measurements on the ensemble of particles are not allowed:
clearly, the results of such experiments can manifest properties of the
composite system of many particles which are not intrinsic properties
of each particle. (We will briefly discuss collective measurements in
Section \ref{swap}.)

After stating here precisely the ``rules of the game'' we now proceed
to discuss the nonlocality of the quantum wave (\ref{qw}) for various
particles.

\section{SINGLE-PHOTON NONLOCALITY}
\label{spnon}

Let us start with considering a photon in a state (\ref{qw}). There
have been several proposals \cite{Tan-sph,Fr-sph,Ha-sph,Ge-sph} how to
obtain quantum correlations based on such and similar systems
\cite{foot1}.  The photon in a state (\ref{qw}) exhibits nonlocality of
the EPR correlations described in Section \ref{corr}.  The state of
the photon, if we write it in the form (\ref{qw2}), is isomorphic to
the EPR state (\ref{sing}). 

In order to get the EPR-type correlations we must be able to perform
measurements on the photon analogous to the spin measurements in
arbitrary direction. The analog of the spin measurement in the $\hat
z$ direction is trivial: it is observing the presence of the photon in
a particular location.  A gedanken experiment yielding the analog of
the spin measurements on the EPR pair in arbitrary directions is as
follows \cite{V-sph}.  Let us consider, in addition to the photon, a
pair of spin$-{1\over 2}$ particles, one located in A and one in B;
see Fig. 3. Both particles are originally in a spin ``down'' state in
the $\hat z$ direction. In the locations A and B there are magnetic
fields in the $\hat z$ direction such that the energy difference
between the ``up'' and ``down'' states equals exactly the energy of
the photon.  Then we construct a physical mechanism of absorption and
emission of the photon by the spin which is described by the unitary
transformation in each site:

 \begin{eqnarray}
 \nonumber
|1\rangle |{\downarrow}\rangle \leftrightarrow |0\rangle
 |{\uparrow}\rangle,\\
 |1\rangle |{\uparrow}\rangle \leftrightarrow |1\rangle
 |{\uparrow}\rangle,\\
\nonumber |0\rangle |{\downarrow}\rangle \leftrightarrow |0\rangle
 |{\downarrow}\rangle.
 \end{eqnarray}

\begin{center} \leavevmode \epsfbox{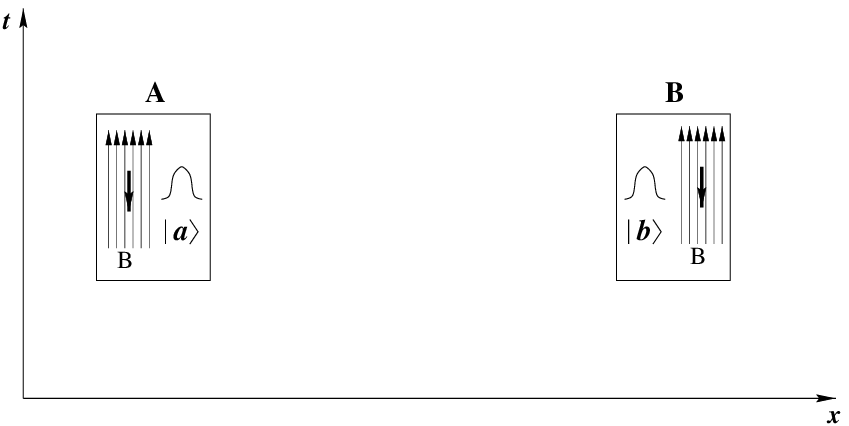} \end{center}

\noindent 
{\small {\bf Fig. 3.} Swapping of the single-photon state with the
  entangled state of two spin$-{1\over 2}$ particles.}

\vskip .8cm
\noindent
This transformation swaps the quantum state of the photon and the
quantum state of the pair of    spin$-{1\over 2}$ particles as follows:
 \begin{eqnarray}
\label{swap1} 
\nonumber
{1\over \sqrt 2}\ (  | 1\rangle_A |0\rangle_B +
e^{i\phi} \,| 0 \rangle_A |1\rangle_B ) \ |{\downarrow}\rangle_A
|{\downarrow}\rangle_B  \rightarrow
~~~~~ \\{1\over \sqrt 2}\  | 0\rangle_A |0\rangle_B \ (  |{\uparrow}\rangle_A
|{\downarrow}\rangle_B   +
e^{i\phi} |{\downarrow}\rangle_A  |{\uparrow}\rangle_B)
   .
\end{eqnarray}
Thus, we can obtain nonlocal correlations of the EPR state starting
with a single  photon, swapping its state to the state of the pair of  spin$-{1\over 2}$ particles,
 and then making appropriate spin component
measurements. Statistical analysis of the correlations between the
results of spin measurements in A and in B 
 allows us to find the phase $\phi$. For example, the probabilities for
 coincidence and anti-coincidence in the $x$ spin measurements are
 given by 
 \begin{eqnarray}
\label{prob-spin}
{\rm prob} (  |{\uparrow}_x \rangle  |{\uparrow}_x  \rangle) =
 {\rm prob} ( | {\downarrow}_x \rangle
|{\downarrow}_x \rangle) = {1\over 4} |1 + e^{i\phi}|^2 ,\\
{\rm prob} ( | {\uparrow}_x \rangle |{\downarrow}_x \rangle) =
 {\rm prob} ( |{\downarrow}_x \rangle
| {\uparrow}_x \rangle) ={1\over 4} |1 - e^{i\phi}|^2 .
\end{eqnarray}

We have shown that, in principal, the nonlocality of a single photon
is equivalent to the nonlocality of the EPR pair. Now we will turn to
the discussion of the possibilities of manifestation of this
nonlocality  in  real experiments and will try to explore
the nature of this equivalence.

We are not aware of experiments in which a spin in a magnetic field
absorbs a photon with high efficiency. However, there is an equivalent
operation which is performed in laboratories. Recently there has been 
a very significant progress in  microwave cavity technology and there are
experiments in which Rydberg atoms which operate as two-level systems
 absorb and emit photons into a microwave
cavity with a very high efficiency \cite{Har}.
The excited state $|e\rangle$ and the ground state $|g\rangle$ of the
atom are isomorphic to $
|{\uparrow}\rangle$ and $ |{\downarrow}\rangle$ states of a
spin$-{1\over 2}$ particle. 
 For the atom, measuring the analog
of the $z$ spin component is trivial: 
 it is the test whether the atom is in the excited state or the ground
state.
 For measurements analogous to the spin
measurements in other directions there is an experimental solution too.
Using appropriate laser pulses the atom state can be ``rotated'' in the
two dimensional Hilbert space of ground and excited states in any
desired way. Thus, any two orthogonal states can be rotated to the
$|e\rangle$ and $|g\rangle$ states and, then, a measurement which
distinguishes between the ground and excited states distinguishes, in
fact, between the original orthogonal states.

The Hamiltonian which leads to the required interactions can be
written in the  following form:
\begin{equation}
  \label{ham}
  H = a^\dagger |g\rangle \langle e| + a  |e\rangle \langle g| ,
\end{equation}
 where $ a^\dagger$, $a$ are creation and annihilation operators of the
 photon. This Hamiltonian is responsible for the two needed operations. First,
 such coupling between the photon in the cavity in A and the atom in A
 together with similar coupling in B swaps the state (\ref{qw2}) to
 the state of two Rydberg atoms:
 \begin{eqnarray}
\label{swap2} 
\nonumber
{1\over \sqrt 2}\ (  | 1\rangle_A |0\rangle_B +
e^{i\phi} \,| 0 \rangle_A |1\rangle_B )\ |g\rangle_A
|g\rangle_B  \rightarrow
~~~~~ \\{1\over \sqrt 2}\  | 0\rangle_A |0\rangle_B \ (  | e\rangle_A |g\rangle_B +
e^{i\phi} | g \rangle_A |e\rangle_B)  
  .
\end{eqnarray}
The same Hamiltonian can also lead to an arbitrary  rotation of
the atomic state. To this end the atom has to be coupled to a cavity
with a {\em coherent state} of  photons, 
\begin{equation}
  \label{alp}
  |\alpha\rangle = e^{-{|\alpha|^2 \over 2}} 
\sum_{n=0}^{\infty} {{\alpha ^n} \over  \sqrt {n!}} \,|n\rangle .
\end{equation}
The phase of $\alpha$  specifies
the axis of rotation and the absolute value of $\alpha$ specifies the rate
of rotation. For example, the time evolution of  an atom starting at $t=0$ in the ground
state is:
 \begin{equation}
\label{psit}
|\Psi (t) \rangle = 
\cos (|\alpha| t) \, |g\rangle + {\alpha \over {i |\alpha|}} \sin
    (|\alpha| t)\, |e\rangle .
\end{equation}
This is correct when we make the approximation $a^\dagger
|\alpha\rangle \simeq \alpha ^\ast |\alpha\rangle$ which is precise in
the limit of large $|\alpha|$.  The Hamiltonian (\ref{ham}) is
actually implemented in laser-aided manipulations of Rydberg atoms
passing through microwave cavities.

Conceptually, the above scheme can be applied to any type of bosons
(instead of photons), even charged bosons.
An example of a (gedanken) Hamiltonian for this case describes a proton $ |p\rangle $ which creates
a neutron $|n\rangle$ by absorbing  a negatively charged meson:
\begin{equation}
  \label{ham-me}
  H = a_m^\dagger |p\rangle \langle n| + a_m  |n\rangle \langle p| ,
\end{equation}
 where $ a_m^\dagger$, $a_m$ are creation and annihilation operators of the
 meson. This Hamiltonian swaps the state  of the meson (now
 written in the form (\ref{qw2})) and the
 state of the nucleon pair:
 \begin{eqnarray}
\label{swap3} 
\nonumber
{1\over \sqrt 2}\ (  | 1\rangle_A |0\rangle_B +
e^{i\phi} \,| 0 \rangle_A |1\rangle_B )\ |p\rangle_A
|p\rangle_B  \rightarrow
~~~~~ \\{1\over \sqrt 2}\  | 0\rangle_A |0\rangle_B \ (  |n\rangle_A |p\rangle_B +
e^{i\phi} |p\rangle_A |n\rangle_B)  
  .
\end{eqnarray}
Since there is no direct measurement of a superposition of proton and
neutron, we need again a procedure which rotates the superposition
states of a nucleon to neutron or proton state. This rotation requires
coherent states of mesons which would be, in this case, a coherent
superposition of states with different charge. Due to strong
electro-magnetic interaction the coherent state will decohere very
fast. This is essentially an environmentally induced ``charge
super-selection rule'' which prevents stable coherent superpositions
of states with different charge. It is important that there is no
{\em exact} charge super-selection rule which would prevent, in principle,
performing the experimental scheme presented above.  Indeed, Aharonov and
Susskind (AS) \cite{AS} proposed a method for  measuring the relative phase
between states with different charge, thus showing that there is no
exact charge super-selection rule. In their method one can measure the
phase even if the whole system (the observed particle and the
measuring device) is in an eigenstate of charge. This 
corresponds to initial entanglement between measuring devices in A and B and
thus will not be suitable for the present procedure. Here we assume
existence of superpositions of different charge states: only then it
is possible that the quantum state of measuring devices in A and B is
a product state.

There are some arguments that the total charge of the universe
is zero and therefore, we cannot have a product of coherent states of
charged particles in A and in B. More sophisticated analysis has to be
performed: since the observable variables are only relative variables,
the final conclusion will be as in the AS paper \cite{AS}:
conceptually, there is no constraint on a measurement of the relative
phase of a charged boson, but decoherence will prevent construction of
any realistic experiment. See also very different arguments against
exact super-selection rule by Giulini \cite{Giu}.

\section{NONLOCALITY OF A FERMION QUANTUM WAVE}
\label{ferm}

As we have shown above, the nonlocality properties of the boson
quantum state (\ref{qw}) are equivalent to the nonlocality  of
the EPR pair. In contrast, the nonlocality properties of the fermion
quantum state (\ref{qw}) are very different from those of the EPR
pair. We cannot generate quantum correlations between results of local
measurements performed in A and in B, the correlations which violate
Bell inequalities.

The reason why the method which was applicable to bosons fails for
fermions is that there is no coherent state of fermions. The number
state $|n\rangle$ exists only for $n=0$ and $n=1$ \cite{foot0}.

The intuitive understanding of the role of the coherent state is as
follows. If, in addition to the measuring devices, there is an
auxiliary identical  particle in a {\em known}
superposition of localized wave packets in A and B, then the phase $\phi$  can be found using local
measurements. 
We consider the superposition of    $ |a'\rangle$ and $ |b'\rangle$
positioned near  $ |a\rangle$ and  $ |b\rangle$ respectively; see
Fig. 4. We choose  the phase of the auxiliary particle to be
equal zero,
 \begin{equation}
\label{qw'}
|\Psi' \rangle = 
{1\over \sqrt 2} (    | a'\rangle +
 | b' \rangle),
\end{equation}
i.e., we have a composite system of two  identical particles in the state
 \begin{equation}
\label{qwqw'}
|\Psi \rangle |\Psi' \rangle = 
{1\over  2} (    | a\rangle +
e^{i\phi} | b \rangle)\, (  | a'\rangle +
 | b' \rangle).
\end{equation}

\begin{center} \leavevmode \epsfbox{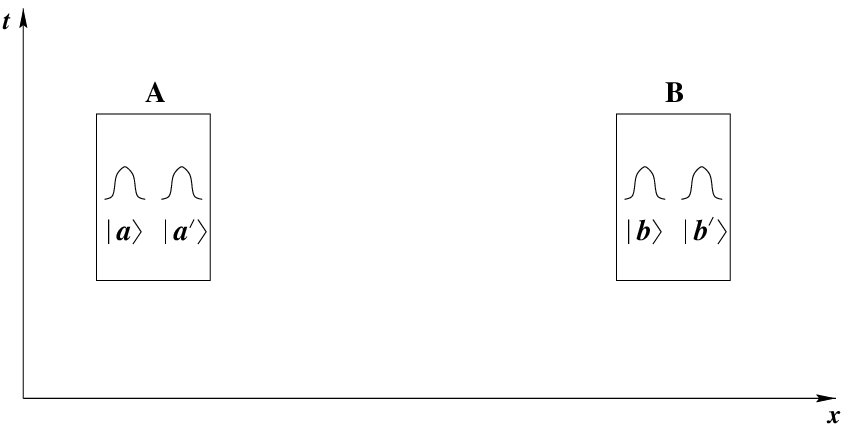} \end{center}

\noindent 
{\small {\bf Fig. 4.} Space-time diagram of local  measurements which
  allow finding the phase $\phi$ of  a quantum wave when an auxiliary
  identical particle with known phase is given.}

\vskip .8cm

 \noindent
 The phase  $\phi$ controls the rate of coincidence counting in
the  measurements of a local  variable in A with eigenstates
 \begin{equation}
\label{qw+}
|a_+ \rangle \equiv
{1\over \sqrt 2} (  | a\rangle +
 | a' \rangle ), ~~~~~~|a_- \rangle \equiv
{1\over \sqrt 2} (  | a\rangle -
 | a' \rangle ) ,
 \end{equation}
and  a local variable in B with eigenstates 
 \begin{equation}
   \label{qw+b}
|b_+ \rangle \equiv
{1\over \sqrt 2} (  | b\rangle +
 | b' \rangle ), ~~~~~~|b_- \rangle \equiv
{1\over \sqrt 2} (  | b\rangle -
 | b' \rangle ).
\end{equation}
In the case that one particle was found on each side, the probabilities are
(compare with (\ref{prob-spin})):
 \begin{eqnarray}
\label{prob1}
{\rm prob} ( |a_+ \rangle |b_+ \rangle) = {\rm prob} ( |a_- \rangle
|b_- \rangle) ={1\over 4} |1 + e^{i\phi}|^2 ,\\
{\rm prob} ( |a_+ \rangle |b_- \rangle) = {\rm prob} ( |a_- \rangle
|b_+ \rangle) = {1\over 4} |1 - e^{i\phi}|^2 .
\end{eqnarray}

The method described in the previous paragraph  is applicable both for
bosons and fermions. However, the existence of a particle described by
(\ref{qw'}) as a part of our measuring devices contradicts our
assumption that sites A and B do not possess an entangled physical
system prior to bringing in the test particle.  For bosons we can
consider a coherent state of particles described by state (\ref{qw'});
it is equal to the product of local coherent states of bosons in A and
in B:
\begin{eqnarray}
  \label{cohe}
\nonumber
   e^{-{|\alpha|^2 \over 2}} 
\sum_{n=0}^{\infty} {{\alpha ^n} \over  \sqrt {n!}} \, {1\over \sqrt
  2^n} (  | a'\rangle + | b' \rangle )^n =~~~~~~~~~~~~~~~~~~~~~~~~\\
 e^{-{|\alpha|^2 \over 2}} 
\sum_{n=0}^{\infty} {{\alpha ^n} \over  \sqrt {n!}} | a' \rangle \  e^{-{|\alpha|^2 \over 2}} 
\sum_{n=0}^{\infty} {{\alpha ^n} \over  \sqrt {n!}} | b' \rangle 
.
\end{eqnarray}
 Thus, this state  has no entanglement between the sites but it
provides the reference  for measuring the phase $\phi$ of the
state (\ref{qw}) via methods described in the previous section.

Again, if we assume that there is no prior entanglement between the
sites A and B, the phase $\phi$ of the fermion quantum state
(\ref{qw}) cannot be measured locally. Quantum correlations which
break Bell's inequality cannot be obtained. The only type of nonlocality
for a fermion wave (except the collapse nonlocality) is the AB
nonlocality. The quantum phase  manifests itself only  in the interference
experiments in which the wave packets $|a\rangle$ and $|b\rangle$ are
brought together.

The impossibility of local measurement of the phase $\phi$ is due to
anti-commutation of fermion operators: the operator $a^\dagger_A +
a_A$ does not commute with the operator $a^\dagger_B + a_B$.  The
eigenstates of the operator $a^\dagger + a$ are ${1\over \sqrt 2}
(|0\rangle \pm |1\rangle)$; we have used measurements of such operator
for finding out the phase $\phi$ of the boson wave in Section
\ref{spnon}.  A measurement in site A of $a^\dagger_A + a_A$ leads to
an observable change in the results of measurement of $a^\dagger_B +
a_B$, where $ a^\dagger_A$, $a_A$, $ a^\dagger_B$, $a_B$ are creation
and annihilation operators of the fermion in A and in B, respectively.
This means that the possibility of such measurements would lead to
superluminal communication.

Another question which can be asked is: ``Can we measure locally the
phase $\phi$ of a superposition of a pair of fermions?'' The quantum
state is:
 \begin{equation}
\label{qw2-2}
|\Psi \rangle = 
{1\over \sqrt 2}\ (  | 2\rangle_A |0\rangle_B +
~e^{i\phi} \,| 0 \rangle_A |2\rangle_B ) ,
\end{equation}
where, for example, $|2\rangle_A$ might represent two electrons in
identical spatial state inside A being in a singlet spin state. Since
$a^\dagger_{\uparrow A} a^\dagger_{\downarrow A} + a_{\uparrow A}
a_{\downarrow A}$ commutes with $a^\dagger_{\uparrow B}
a^\dagger_{\downarrow B} + a_{\uparrow B} a_{\downarrow B}$, the
argument presented in the preceding paragraph for unmeasurability of
the phase of a superposition of  single-fermion wave packets does not
hold in this case. In fact, a pair of fermions is, in a sense, a
boson. We can construct a procedure for measuring phase $\phi$ of the
state (\ref{qw2-2}) similar to the procedure which was previously described 
(for a photon (\ref{ham})-(\ref{psit}) and  for a
charged meson (\ref{ham-me}), (\ref{swap3})) in Section \ref{spnon}.
A difficulty is that the coherent state of pairs of fermions
which is required for our procedure can only be constructed
approximately.

\section{IS IT POSSIBLE TO CHANGE THE PHASE IN A NONLOCAL WAY?}
\label{disc}

The main message of Section \ref{spnon} is that the phase $\phi$ for
boson state (\ref{qw}) is locally measurable. Given an ensemble of
bosons with identical phase $\phi$ we can generate a set of numbers
(results of measurements) in A and another set of numbers in B, such that
the two sets together yield $\phi$. This sounds paradoxical, in particular,
because  $\phi$ is not a gauge invariant parameter.

Moreover, it seems that this phase can be changed non-locally. Indeed,
it has been described in Section \ref{ferm} how opening a condenser
for a period of time in the space between the locations of a charged
particle, A and B, changes the phase: this is the scalar AB effect.
Thus, it seems that by an action in a localized region we can send
information to a space-like separated region. Opening or not opening a
condenser apparently changes correlations in the results of
measurements in A and B; see Fig. 5.

\begin{center} \leavevmode \epsfbox{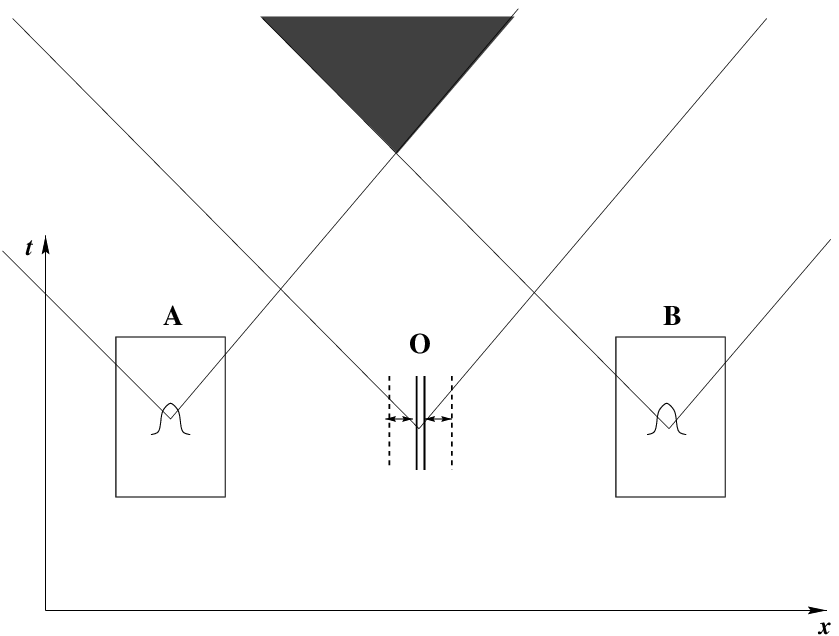} \end{center}

\noindent 
{\small {\bf Fig. 5.} Apparent sending signals to a space-like
  separated region. Operation in O, opening the condenser for a period
  of time, apparently changes the correlations between measurements in
  A and B. No signal is sent from O, neither to A nor to B, but the
  signal {\em is} sent to the union of A and B. The intersection of
  light cones originated at A and at B lies inside the light cone
  originated at O.  Therefore, the action of the condenser falls into
  the category of ``jammers'' considered in Ref.  \cite{GPR}.}

\vskip .6cm

It has been shown \cite{GPR}
  that such an action, if possible, cannot lead to a paradoxical
  causal loop similar to the one generated by a possibility of sending
  signals from one localized space-time region to another space-like
  separated {\em local} region.  In our case the region to which we
  send the information consists  of two space-like separated
  regions.  There is no local observer who receives superluminal
  signals.

  In spite of the fact that we cannot reach a causality paradox if
  such operation is possible, it  clearly contradicts the spirit, if
  not the letter, of special relativity. And, in fact, it is
  impossible. It is incorrect that the opening of a condenser will change
  correlations between results of measurements in A and B. It must be
  incorrect because we should be able to use a covariant gauge in
  which changes in the potentials take place only inside the light
  cone. However, we can explain this phenomena also in a standard
  (Coulomb) gauge. In our
  scheme the measurements in local sites include interactions with
  coherent states of auxiliary particles,  particles which are
  identical to the particle in a superposition. Therefore, if the
  particle in question is charged, the auxiliary particles are also charged
  and opening the condenser  changes the phase of the coherent
  state in such a way that the correlations are  not  changed. The
  gauge which we choose changes the description of auxiliary particles
  too, so that the probabilities for results of measurements remain
  gauge invariant.
  
  Consider now a neutral boson state. A massive plate in between the
  regions A and B which we move or not move toward one of the sites
  will introduce the phase shift in complete analogy with the scalar
  AB effect. (The difference here is that the gravitational fields in
  the regions A and B are not zero, but the fields are not affected by
  the motion of the plate.) In a scenario where the boson is absorbed
  by spins in a magnetic field and the correlations are obtained from
  the spin measurements, it is not obvious how the measuring devices
  will be influenced by the movement of the massive plate.  The
  resolution of the paradox in this case is similar to the resolution
  of  Einstein's paradox of an exact energy of an exact clock
  \cite{ETP}.  The explanation is that the pointers of the local
  clocks are shifted.  Simultaneity between A and B is altered due to
  the action of the massive plate.  Since in our case local clocks
  activate the measurements, the shift in the pointer will lead to a
  change. This change compensates exactly the phase change of the
  boson.

\section{COLLECTIVE MEASUREMENTS}

\label{swap}

In this paper we have considered the results of measurements on an
ensemble of identical particles in an unknown state. We allow
measurements to be performed only on  single members of the ensemble,
so that we will have an ensemble of results of measurements performed on single
particles. We believe that this is the proper approach for the
analysis of the nature of a quantum wave of a particle; however,  it might
be interesting to consider a related question: ``Are there any changes
to the questions posed in this paper if collective measurements are
allowed?'' Note that there is a recent result showing that collective
measurements do make a difference for similar questions regarding the
nonlocality of an ensemble of pairs of spin-$1\over 2$ particles in a
particular mixed state \cite{Peres2}.

For bosons we do not expect any difference because, even for
single-particle measurements, we got the answers to our questions:
(i) statistical analysis of the results of measurements allows us to find
 the phase $\phi$ and (ii) there are measurements  in A and in B
such that the results are characterized by
correlations which can not have local causes. 
 For single-particle measurements on fermions both (i) and (ii) are not
 true and thus raises  the  question of the status of (i) and (ii) when
 collective measurements are allowed.
 
 Let us start this analysis by assuming that our particle is an
electron and, contrary to the assumption of no prior entanglement, we
 now  have an auxiliary particle, a positron, in a known superposition
  in A and B, say, of the form (\ref{qw'}). In this case both
 (i) and (ii) are true: the fermion state {\em is} measurable via
 local measurements, and some measurements in A and B exhibit
 correlations which have no local causes.

 Indeed, we can apply an interaction such that the positron and the
 electron located in the same site annihilate and create a photon.
 Such interaction will lead to the following transformation
\begin{eqnarray}
\label{anni} 
\nonumber
{1\over \sqrt 2} (  | e^-\rangle_A  +
e^{i\phi} | e^- \rangle_B)\ {1\over \sqrt 2}( |e^+\rangle_A  +
|e^+\rangle_B)
 \rightarrow ~~~~~\\
{1\over  2} (  |e^-\rangle_A |e^+\rangle_B +e^{i\phi} | e^-
 \rangle_B |e^+\rangle_A + | \gamma\rangle_A  + e^{i\phi} | \gamma
 \rangle_B ) .
\end{eqnarray}
After testing and {\em not} finding the electron and the positron in the sites A and
B the remaining state will be \cite{foot}:
\begin{equation}
\label{phot} 
 {1\over \sqrt 2} (  | \gamma\rangle_A  + ~e^{i\phi} | \gamma \rangle_B ),
\end{equation}
which is a different notation for a single-photon state of the form (\ref{qw}). For a single photon we know that (i) and (ii) are true: the phase of a
single-photon state (which is the original  phase $\phi$ of
the fermion)  can be found, and quantum correlations breaking Bell
inequalities can be obtained.

However, we do not have a positron in a state (\ref{qw'}). Instead, we
have an ensemble of electrons in a state (\ref{qw}). So, the first
step is to {\em swap} the state of the electron with the state of a
positron \cite{foot2}.  If we have an entangled state of a composite
system which has two parts, one in A and another in B, such as the EPR
state of two spin-${1\over 2}$ particles located in A and B, and we
want to transfer this entangled state to another pair of particles in
A and B, then all we have to do is to perform local operation in each
site which swaps the local quantum states of one particle from one
pair with one particle from the other pair located in the same site
\cite{AAV86}. Linearity of quantum mechanics will ensure that swapping
of local states, i.e. the states of parts of the systems, will lead to
swapping of the quantum state of the whole systems.
 
In this paper we are interested in the swapping of a nonlocal state of
a single particle to another single particle. The method described
above cannot be applied directly because it is assumed that we have no
another particle in a superposition of being in A and B (this is
entanglement). Therefore, the other particle is not present in at
least one of the sites and consequently, the ``local swapping
interaction'' with this particle is meaningless. However, if the
particles are bosons, then the swapping operation is possible.  It can
be done by transferring the quantum state to the entangled state of a
composite system: a single-photon state can be transferred to two
spin-${1\over 2}$ particles in a magnetic field in the gedanken
scenario described in Section \ref{spnon} or to two atoms in a real
experiment using microwave cavities. After that, the quantum state can
be swapped back to ``another'' photon.

Let us come back to the question of transferring  the quantum state of
the electron to a positron.  Again, since we assumed no prior
entanglement, the positron cannot be in a superposition of being in A
and in B. Therefore, we will consider a situation in which there are two
positrons one in A and another in B. We apply an interaction such that
the positron and the electron which are in the same site annihilate
and create a photon. This is described by the equation
\begin{eqnarray}
\label{inter} 
\nonumber
{1\over \sqrt 2} (  | e^-\rangle_A  +
e^{i\phi} | e^- \rangle_B) \ |e^+\rangle_A |e^+\rangle_B \rightarrow ~~~~\\
 {1\over \sqrt 2}\ (  | \gamma\rangle_A  |e^+\rangle_B +
e^{i\phi} | \gamma \rangle_B |e^+\rangle_A) .
\end{eqnarray}
Now, the procedure described in Section \ref{spnon} allows
measurements of local superpositions of the vacuum and  single-photon 
states. In particular, there is a nonzero probability to find the state
${1\over \sqrt 2} ( |0\rangle_A + |\gamma \rangle_A)$ in A and a
similar state  ${1\over \sqrt 2} ( |0\rangle_B + |\gamma \rangle_B)$ in B.
When this occurs, the final situation is that the electron and one of
the positrons are annihilated and a positron appears in a superposition
of being in two places
\begin{equation}
\label{posit} 
 {1\over \sqrt 2} (   |e^+\rangle_B +
~e^{i\phi}   |e^+\rangle_A) .
\end{equation}
Thus, we can obtain a positron in superposition from an electron in a
superposition. If we are allowed to perform collective measurements we
now can annihilate this positron with another electron in the ensemble:
\begin{eqnarray}
\label{anni2} 
\nonumber
{1\over \sqrt 2} (  | e^-\rangle_A  + e^{i\phi} | e^- \rangle_B)\ 
{1\over \sqrt 2}( ~e^{i\phi}  |e^+\rangle_A  + |e^+\rangle_B)
 \rightarrow ~~~~\\
 {1\over  2} (  |e^-\rangle_A |e^+\rangle_B +e^{i2\phi} | e^-
\rangle_B |e^+\rangle_A +e^{i\phi}  | \gamma\rangle_A  + 
e^{i\phi} | \gamma \rangle_B )
\end{eqnarray}
We do not obtain relative phase between photon wave-packets in two
places which would allow us to find the  phase $\phi$, but  we do obtain a
superposition of a photon in A and B with known (zero) phase. This
superposition can generate quantum correlations without local
causes as described above.  

If we are allowed to perform collective measurements, we can consider
measurements on the pairs of fermions from our ensemble. The phase of
pairs of fermions is $2\phi$ and, in general, it can be found by the
method described in Section \ref{spnon}. However, as we mentioned
above, all statements about measurability using collective
measurements do not describe the nature of a quantum wave of a single
particle.

\section{CONCLUSIONS}

\label{conc}

In this paper we have analyzed nonlocal aspects of a simple quantum
wave which is an equal-weights superposition (\ref{qw}) of wave packets
in A and in B.  For this analysis we assumed that we are given
non-entangled laboratories in A and B 
which are described quantum mechanically by a product state of systems
in A and systems in B.

We have shown that presence of an ensemble of bosons in a
superposition ${1\over \sqrt 2} ( | a\rangle + e^{i\phi} | b \rangle)$
leads to correlations in the results of single-particle local
measurements in A and in B which break Bell's inequality.  These
results, collected from a large ensemble allows us to find the
phase $\phi$. Thus, the boson quantum wave exhibit the EPR-type
nonlocality. For a photon state this is not just a theoretical
statement: the EPR nonlocality can be observed in an ensemble of
measurements carried out on single photons. In principle, the
statement applies to  any boson state.  However, environmentally
induced super-selection rule prevents such experiment with charged
bosons. Also, experiments with neutral massive bosons do not seem to
be feasible.
 
The presence of an ensemble of fermions in a superposition ${1\over
  \sqrt 2} ( | a\rangle + e^{i\phi} | b \rangle)$ with the restriction
that we perform separate measurements on each fermion does not lead to
correlations in the results of the local measurements in A and in B
which violate Bell's inequality. We do get correlations between the
results of local measurements in A and B, but these correlations are
of the kind which allow local causal explanation.  These results do
not allow us to find the phase $\phi$. The phase $\phi$ has observable
consequences in interference experiments. A fermion quantum wave
exhibits the AB nonlocality which is the  unobservability of this phase
via local single-particle measurements.

\vspace{.3cm}
 \centerline{\bf  ACKNOWLEDGMENTS}
 
 It is a pleasure to thank Lior Goldenberg, Jacob Grunhaus, Benni
 Reznik, Sandu Popescu, Asher Peres and especially Philip Pearle for
 helpful discussions.  This research was supported in part by grant
 471/98 of the Basic Research Foundation (administered by the Israel
 Academy of Sciences and Humanities) and NSF grant PHY9601280.  Part
 of this work was done during the 1999 ESF-Newton Institute Conference
 in Cambridge.

\end{multicols}


\begin{thebibliography}{99}
\bibitem{foot0} Although we have not mentioned ``coherent state'' in
  the gedanken experiment of finding the phase $\phi$ of a single
  photon state through swapping it to the state of the pair of
  spin$-{1\over 2}$ particles, the analog of the coherent state was
  present in that case too. The ability of measuring a spin component
  in arbitrary direction requires a reference frame for the direction
  in space. The quantum state with a well defined angle is a
  superposition of different angular momentum states which is
  analogous to the superposition of different number states in the
  coherent state of photons.
\end{thebibliography}
\end{document}